\documentclass{elsart}
\usepackage{epsfig}

\begin{document}
\begin{frontmatter}
\title{Measurement of the Neutron Spin Structure Function 
g$_{2}^{n}$ and Asymmetry A$_{2}^{n}$}
\collab{E154 Collaboration}
\author[19]{K. Abe}, 	  	
\author[16]{T. Akagi},
\author[8]{B. D. Anderson},
\author[16]{P. L. Anthony},
\author[1]{R. G. Arnold},
\author[6]{T. Averett},
\author[21]{H. R. Band},
\author[9]{C. M. Berisso},
\author[14]{P. Bogorad},
\author[7]{H. Borel},
\author[1]{P. E. Bosted},
\author[2]{V. Breton},
\author[16]{M. J. Buenerd\thanksref{AA}},
\author[14]{G. D. Cates},
\author[10]{T. E. Chupp},
\author[9]{S. Churchwell},
\author[10]{K. P. Coulter},
\author[16]{M. Daoudi},
\author[15]{P. Decowski},
\author[16]{R. Erickson},
\author[1]{J. N. Fellbaum},
\author[2]{H. Fonvieille},
\author[16]{R. Gearhart},
\author[5]{V. Ghazikhanian},
\author[20]{K. A. Griffioen},
\author[9]{R. S. Hicks},
\author[17]{R. Holmes},
\author[6]{E. W. Hughes},
\author[5]{G. Igo},
\author[2]{S. Incerti},
\author[21]{J. R. Johnson},
\author[17]{W. Kahl},
\author[8]{M. Khayat},
\author[9]{Yu. G. Kolomensky},
\author[12]{S. E. Kuhn},
\author[14]{K. Kumar},
\author[19]{M. Kuriki},
\author[7]{R. Lombard-Nelsen},
\author[8]{D. M. Manley},
\author[7]{J. Marroncle},
\author[16]{T. Maruyama},
\author[15.5]{T. Marvin},
\author[16]{W. Meyer\thanksref{BB}},
\author[18]{Z.-E. Meziani},
\author[11.5]{D. Miller},
\author[21]{G. Mitchell},
\author[8]{M. Olson},
\author[9]{G. A. Peterson},
\author[8]{G. G. Petratos},
\author[16]{R. Pitthan},
\author[21]{R. Prepost},
\author[13]{P. Raines},
\author[12]{B. A. Raue\thanksref{CC}},
\author[1]{D. Reyna},
\author[16]{L. S. Rochester},
\author[1]{S. E. Rock},
\author[14]{M. V. Romalis},
\author[7]{F. Sabatie},
\author[4]{G. Shapiro},
\author[9]{J. Shaw},
\author[10]{T. B. Smith},
\author[1]{L. Sorrell},
\author[17]{P. A. Souder},
\author[7]{F. Staley},
\author[16]{S. St. Lorant},
\author[16]{L. M. Stuart},
\author[19]{F. Suekane},
\author[1]{Z. M. Szalata},
\author[7]{Y. Terrien},
\author[11]{A. K. Thompson},
\author[1]{T. Toole},
\author[17]{X. Wang},
\author[8]{J. W. Watson},
\author[10]{R. C. Welsh},
\author[12]{F. R. Wesselmann},
\author[21]{T. Wright},
\author[16]{C. C. Young},
\author[16]{B. Youngman},
\author[19]{H. Yuta},
\author[8]{W.-M. Zhang} and
\author[18]{P. Zyla}
\address[1]{The American University, Washington D.C. 20016}
\address[2]{Universit\'{e} Blaise Pascal, LPC IN2P3/CNRS, F-63170
Aubi\`{e}re Cedex, France} 
\address[4]{University of California, Berkeley,
California 94720-7300}
\address[5]{University of California, Los Angeles,
California  90024} 
\address[6]{California Institute of Technology, Pasadena, California 91125}
\address[7]{Centre d'Etudes de Saclay, DAPNIA/SPhN, F-91191
Gif-sur-Yvette, France} 
\address[8]{Kent State University, Kent, Ohio 44242}
\address[9]{University of Massachusetts, Amherst, Massachusetts 01003}
\address[10]{University of Michigan, Ann Arbor, Michigan 48109}
\address[11]{National Institute of Standards and Technology,
Gaithersburg, Maryland 20899}
\address[11.5]{Northwestern University, Evanston, Illinois 60201} 
\address[12]{Old Dominion University, Norfolk, Virginia  23529}
\address[13]{University of Pennsylvania, Philadelphia, Pennsylvania 19104}
\address[14]{Princeton University, Princeton, New Jersey 08544}
\address[15]{Smith College, Northampton, Massachusetts  01063}
\address[15.5]{Southern Oregon State College, Ashland, Oregon  97520}
\address[16]{Stanford Linear Accelerator Center, Stanford,
California 94309} 
\address[17]{Syracuse University, Syracuse, New York 13210}
\address[18]{Temple University, Philadelphia, Pennsylvania 19122}
\address[19]{Tohoku University, Aramaki Aza Aoba, Sendai, Miyagi, Japan}
\address[20]{College of William and Mary, Williamsburg, Virginia 23187}
\address[21]{University of Wisconsin, Madison, Wisconsin 53706}
\thanks[AA]{Permanent Address: Institut des Sciences Nucl\'{e}aires,
IN2P3/CNRS, 38026 Grenoble Cedex, France}
\thanks[BB]{Permanent Address: University of Bochum, D-44780 Bochum, Germany}
\thanks[CC]{Present Address: Florida International
University, Miami, FL 33199}
 
\begin{abstract}
We have measured the neutron structure function g$_{2}^{n}$ and the
virtual photon-nucleon asymmetry A$_{2}^{n}$ over the kinematic range
$0.014\leq x \leq 0.7$ and $1.0 \leq Q^{2} \leq 17.0$ by scattering 48.3
GeV longitudinally polarized electrons from polarized $^{3}$He.
Results for A$_{2}^{n}$ are significantly smaller than
the $\sqrt{R}$ positivity limit over most of the measured range and
data for g$_2^{n}$ are generally consistent with the twist-2 
Wandzura-Wilczek prediction.
Using our measured g$_{2}^{n}$ we obtain results
for the twist-3 reduced matrix element $d_{2}^{n}$, and the
integral $\int$g$_{2}^{n}(x)dx$ in the range $0.014\leq x \leq 1.0$.  
Data from this experiment are combined with existing data for g$_{2}^{n}$ 
to obtain an average for $d_{2}^{n}$ and the integral
$\int$g$_{2}^{n}(x)dx$.
\end{abstract}
\end{frontmatter}

%PACS numbers 13.60.Hb, 13.88.+e, 24.70.+s, 25.30.Fj
%Keywords: nucleon, spin structure, deep inelastic scattering, 
%          neutron, spin, structure function, transverse

\newpage
The deep inelastic spin structure functions g$_{1}(x,Q^{2})$ and 
g$_{2}(x,Q^{2})$, which depend on the Bjorken scaling variable $x$ and
the virtual photon four-momentum squared $-Q^{2}$, provide insight
into the internal spin structure of the nucleon.
A large set of data for g$_{1}$ now exists 
for the proton, deuteron~\cite{SMC,E143} and
neutron~\cite{E142_2,E154}.  These data have been used
to test the fundamental Bjorken sum rule, and 
within the framework of the quark-parton model (QPM), to
measure the quark contribution to the nucleon's spin. 
The g$_{2}$ structure function contains contributions from both the 
longitudinal and transverse polarization distributions within the
nucleon.  It is sensitive to higher twist effects such as quark-gluon
correlations and quark mass contributions, and is not easily
interpreted in the QPM where such effects are not included.
However, by interpreting g$_{2}$ using the operator product expansion (OPE)
within QCD~\cite{Vain,Jaffe}, it is possible to study contributions to
the nucleon spin structure beyond the simple QPM.

The OPE allows us to write the hadronic matrix element in deep 
inelastic scattering (DIS) in terms of a series of renormalized operators of
increasing twist~\cite{Vain,Jaffe}.  
The leading contribution is twist-2, with higher
twist terms suppressed by powers of $1/Q$.
Keeping only terms up to twist-3, the moments of g$_{1}$ and g$_{2}$
at fixed $Q^{2}$ can be related to the twist-2 and twist-3
reduced matrix elements, $a_{j}$ and $d_{j}$~\cite{Jaffe},
\begin{eqnarray}
\label{eq:moments}
\int_{0}^{1}x^{j}{\rm{g}}_{1}(x,Q^{2})dx=\frac{a_{j}}{2},
\;\;j=0,2,4,... \nonumber \\ 
\int_{0}^{1}x^{j}{\rm{g}}_{2}(x,Q^{2})dx=\frac{1}{2}\frac{j}{j+1}(d_{j}-a_{j}),
\;\;j=2,4,...
\end{eqnarray}
In the expressions above, $d_{j}$ directly appears in the equation
for g$_{2}$ allowing us to study the higher twist structure of the 
nucleon at leading order.
An expression for the twist-2 part of g$_{2}$ was derived by 
Wandzura and Wilczek~\cite{g2ww} 
from these sum rules assuming that the twist-3 contributions
$d_{j}$, are negligible,
\begin{equation}
{\rm{g}}_{2}^{WW}(x,Q^{2})=-{\rm{g}}_{1}(x,Q^{2}) + \int_{x}^{1}
\frac{{\rm{g}}_{1}(x',Q^{2})}{x'} dx'.
\label{eq:g2ww}
\end{equation}
Comparing measured values of g$_{2}$ with this prediction enables us
to extract information about higher twist contributions to g$_2$.
There is an additional twist-2
contribution to g$_{2}$~\cite{Song,PCR} beyond the g$_{2}^{WW}$ term 
which arises from the transverse polarization density in the 
nucleon, $h_{T}(x,Q^{2})$.
However, this term is suppressed by the ratio of the quark to nucleon
mass $m/M$ in DIS~\cite{Song} and will be neglected in this analysis.

The structure function
g$_{2}(x,Q^{2})$ may be expressed in terms of two
measurable asymmetries, $A_{\parallel}(x,Q^{2})$ and
$A_{\perp}(x,Q^{2})$,
corresponding to longitudinal and transverse target polarization
with respect to the incoming electron beam helicity, 
\begin{equation}
{\rm{g}}_{2}(x,Q^{2})=\frac{F_{2}(x,Q^{2})(1+\gamma ^{2})}
{2x\left[1+R(x,Q^{2})\right]}
\frac{y}{2d\,{\rm{sin}}\, \theta}
\biggl[A_{\perp}\frac{E+E'{\rm{cos}}\, \theta}{E'} 
-A_{\parallel}\: {\rm{sin}}\, \theta \biggr],
\end{equation}
where  $E$  and $E'$ are the incident and scattered electron energies, 
$\theta$ is the scattering angle, $\gamma=2Mx/\sqrt{Q^{2}}$, $y=(E-E')/E$,
$d=(1-\epsilon)(2-y)/y[1+\epsilon R(x,Q^{2})]$, and $\epsilon^{-1}=
1+2\left[1+\gamma ^{-2}\right]{\rm{tan}}^{2}(\theta /2)$.  Fits to existing 
data were used for the unpolarized structure function 
$F_{2}(x,Q^{2})$~\cite{NMC} and for $R(x,Q^{2})$~\cite{R1990}, the ratio of
longitudinal to transverse virtual photon absorption cross sections.
At small scattering angles, the term $A_{\parallel}\:{\rm{sin}}\,\theta$
is small, and consequently the dominant contribution to g$_2$ comes
from $A_{\perp}$.

Spin dependent DIS can also be described in terms of the spin
asymmetries A$_{1}(x,Q^{2})$ and A$_{2}(x,Q^{2})$ for virtual photon
absorption. 
The asymmetry A$_2(x,Q^{2})$ is bounded by the positivity limit
$|{\rm{A}}_{2}(x,Q^{2})|\leq \sqrt{R(x,Q^{2})}$, and like g$_{2}$, it
is dominated by $A_{\perp}$,
\begin{equation}
{\rm{A}}_{2}(x,Q^{2})=\frac{\gamma (2-y)}{2d\, 
{\rm{sin}}\,\theta }\left[A_{\perp}
\frac{y(1+xM/E)}{(1-y)} +  A_{\parallel}\: {\rm{sin}}\,\theta \right].
\end{equation}

Measurements of g$_2$ and A$_{2}$ exist for the
proton~\cite{E143g2,SMCg2} and deuteron~\cite{E143g2}, and in the case
of the neutron, a measurement was made at the Stanford Linear Accelerator
Center (SLAC) by scattering 26 GeV polarized electrons from polarized
$^3$He~\cite{E142_2}.
In this Letter, we report a new measurement of g$_{2}$ and A$_{2}$
for the neutron made during experiment E154 at SLAC.
For this experiment, the beam energy was increased to 48.3 GeV and 
two new large-acceptance spectrometers were constructed to provide
broader kinematic coverage than previously measured.   
Results from this experiment for g$_{1}^{n}$ and A$_{1}^{n}$ have been 
reported elsewhere~\cite{E154}, and we focus here 
on the measurement of $A_{\perp}$ and the subsequent determination of
g$_{2}^{n}$ and A$_{2}^{n}$.  

The target was a 30 cm long, thin-walled glass cell containing
approximately 10.5 atmospheres (as measured at $20^{\circ}$C) 
of $^{3}$He gas.  The helium nuclei were polarized by
spin-exchange collisions with rubidium atoms that were polarized
by optical pumping~\cite{target}.  The system was 
designed to allow continuous pumping of the target polarization in 
the longitudinal direction only.
Therefore, to obtain transverse polarization, the $^3$He spins were
first pumped to a longitudinal polarization of 48\% and
then rotated to the transverse direction using 
two orthogonal sets of Helmholtz coils.
In the transverse orientation, the polarization decreased to 33\%
over a period of 24 hours, at which time the target was 
re-polarized.  Approximately $7\times 10^{6}$ electron 
events were recorded during two cycles of transverse running.

The electron beam was produced in 250 ns pulses
at a rate of 120 Hz, each containing approximately 
$3\times 10^{10}$ electrons.
The average beam polarization was measured to be $0.826 \pm 0.023$
using a M{\o}ller polarimeter~\cite{moller}, and the helicity of
each pulse was chosen randomly to reduce helicity-dependent 
systematic errors.  Scattered electrons were measured using
two independent spectrometers at scattering
angles centered around $2.75 ^{\circ}$ and $5.5^{\circ}$, and the
asymmetry $A_{\perp}$ was calculated as
\begin{equation}
A_{\perp}=\left(\frac{N^{-}-N^{+}}{N^{-}+
N^{+}}\right)\frac{1}{P_{b}P_{t}f},
\end{equation}
where $N^{+}$ and $N^{-}$ are the measured electron rates for positive
and negative beam helicities corrected for detection efficiency and normalized
to the incident charge, $P_{b}$ and $P_{t}$ are the beam and target
polarizations, and  
$f$ is the dilution factor which corrects for electrons that scattered from
materials in the target system other than $^3$He. 

Our measured asymmetry included not only DIS events, but also
pions mis-identified as electrons, and electrons produced in 
charge symmetric hadron
decays.  The rates and asymmetries for these backgrounds were
measured and used to correct $A_{\perp}$.
The asymmetry was also corrected for internal~\cite{Kukhto} and
external~\cite{Tsai} radiative effects.  Uncertainties in the
radiative corrections were estimated by varying the input models over
a range consistent with the measured data.
Finally, a neutron result was extracted from $A_{\perp}$ by applying a
correction for the $^{3}$He nuclear wave function~\cite{he3_1} and
using g$_{2}^{WW}$ obtained
from a fit to existing proton data~\cite{SMC,E143} for g$_{1}^{p}$.

\begin{figure}[t]
{\epsfig{figure=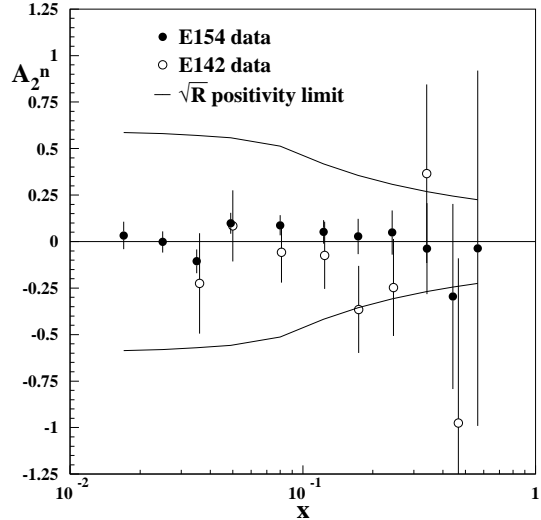,width=7.0cm}}
\caption{The asymmetry A$_{2}^{n}$ 
for both spectrometers combined, and the
corresponding $\protect\sqrt{R}$ positivity limits.  Also
shown are the neutron data from SLAC experiment E142.  Errors are
statistical only.}  
\label{a2}
\end{figure}

\begin{figure}
{\epsfig{figure=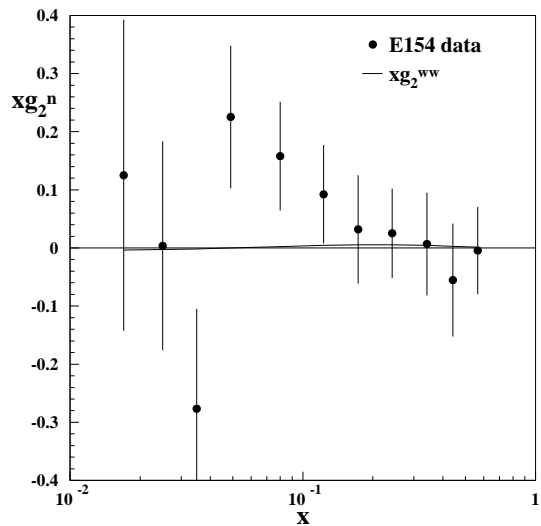,width=7.0cm}}
\caption{The structure function $x$g$_{2}^{n}$ for both 
spectrometers combined.  Also shown is the the twist-2 g$_{2}^{WW}$ 
prediction.  Errors are statistical only.}
\label{xg2}
\end{figure}

Results for A$_{2}^{n}$ and g$_{2}^{n}$ from both spectrometers
are given in Table~\ref{tab:data} with statistical and systematic errors.
The data cover the kinematic range $0.014\leq x \leq 0.7$ and 
$1.0\leq Q^{2} \leq 17.0$ (GeV/c)$^{2}$ with an average $Q^2$ of $3.6$
(GeV/c)$^2$.  Systematic errors are 
dominated by the uncertainty in the radiative corrections, but are
significantly smaller than the statistical error over the entire data range.  
No evidence of $Q^{2}$ dependence was seen for A$_{2}^{n}$ or
g$_{2}^{n}$ within the
experimental errors and the data from both spectrometers were
averaged.  The results for A$_{2}^{n}$ are shown 
in Fig.~\ref{a2} along with the $\sqrt{R}$ positivity limit and previous data
from SLAC experiment E142~\cite{E142_2}.  The data are in good
agreement with the E142 measurement and are significantly smaller than
the positivity limit over most of the measured range.
Results for $x{\rm{g}}_{2}^{n}$ are shown in Fig.~\ref{xg2} along with  
the twist-2 prediction, $x{\rm{g}}_{2}^{WW}$.   To calculate g$_{2}^{WW}$, 
we assume that g$_{1}/F_{1}$ is independent of Q$^2$, and use a fit 
to our measured g$_1^n$ data~\cite{E154}. A comparision of our
data with g$_{2}^{WW}$ over the measured range gives
a $\chi^{2}/$(dof) of $1.02$ indicating good overall 
agreement.  However, the data clearly do not rule out the possibility
of large twist-3 contributions and show marginal agreement with the
twist-2 prediction in the region $0.03<x<0.1$.
\begin{table}[t]
\caption{Results for A$_{2}^{n}$ and g$_{2}^{n}$ for the
$2.75^{\circ}$ and $5.5^{\circ}$ spectrometers.}
\begin{tabular}{rrcrr}
\hline \hline
\multicolumn{1}{c}{$x$ range} & \multicolumn{1}{c}{$<x>$} & 
\multicolumn{1}{c}{$<Q^{2}>$} & \multicolumn{1}{c}{A$_{2}^{n}$} & 
\multicolumn{1}{c}{g$_{2}^{n}$}\\ 
 & & \multicolumn{1}{c}{(GeV/c)$^2$} & 
\multicolumn{1}{c}{$\pm$stat$\pm$syst} & 
\multicolumn{1}{c}{$\pm$stat$\pm$syst} \\
\hline
\multicolumn{5}{c}{$2.75^{\circ}$ Spectrometer} \\ \hline
0.014 - 0.02 & 0.017 & 1.2 & $0.03\pm 0.07\pm  0.01$ & $7.36\pm 
15.74\pm  2.24$\\
0.02  - 0.03 & 0.025 & 1.6 & $0.00\pm  0.06\pm  0.01$ & $0.15\pm
7.19\pm  0.98$\\
0.03  - 0.04 & 0.035 & 2.1 & $-0.11\pm  0.06\pm  0.01$ & $-7.90\pm
4.91\pm  0.96$\\
0.04  - 0.06 & 0.049 & 2.6 & $0.10\pm  0.06\pm  0.01$ & $4.60\pm  
2.50\pm  0.54$\\
0.06  - 0.10 & 0.077 & 3.4 & $0.06\pm  0.06\pm  0.01$ & $1.32\pm  
1.34\pm  0.25$\\
0.10  - 0.15 & 0.122 & 4.1 & $0.13\pm  0.11\pm  0.03$ & $1.22\pm  
0.95\pm  0.24$\\
0.15  - 0.20 & 0.173 & 4.7 & $-0.03\pm  0.18\pm  0.03$ & $-0.08\pm
0.81\pm  0.14$\\
0.20  - 0.30 & 0.242 & 5.1 & $-0.25\pm  0.24\pm  0.05$ & $-0.48\pm
0.51\pm  0.11$\\
0.30  - 0.40 & 0.341 & 5.6 & $0.63\pm  0.55\pm  0.13$ & $0.54\pm  
0.46\pm  0.15$\\
0.40  - 0.50 & 0.425 & 5.9 & $0.16\pm  1.40\pm  0.04$ & $0.04\pm  
0.57\pm  0.02$\\
\hline
\multicolumn{5}{c}{$5.5^{\circ}$ Spectrometer} \\ \hline
0.06  - 0.10 & 0.084 & 5.5 & $0.16\pm  0.10\pm  0.02$ & $4.08\pm  
2.40\pm  0.43$\\
0.10  - 0.15 & 0.123 & 7.2 & $0.01\pm  0.08\pm  0.02$ & $0.23\pm
1.00\pm 0.20$\\
0.15  - 0.20 & 0.173 & 8.9 & $0.05\pm  0.11\pm  0.02$ & $0.40\pm  
0.72\pm  0.15$\\
0.20  - 0.30 & 0.242 & 10.7 & $0.15 \pm 0.14\pm  0.03$ & $0.48\pm
0.41\pm  0.10$\\
0.30  - 0.40 & 0.342 & 12.5 & $-0.21\pm  0.27\pm  0.03$ &
$-0.22\pm 0.31\pm  0.04$\\
0.40  - 0.50 & 0.442 & 13.8 & $-0.36\pm  0.53\pm  0.05$ &
$-0.16\pm  0.24\pm  0.03$\\
0.50  - 0.70 & 0.564 & 15.0 & $-0.04\pm  0.96\pm  0.06$ &
$-0.01\pm  0.13\pm  0.01$\\
\hline
\hline
\end{tabular}
\label{tab:data}
\end{table}

In order to quantify the possible contribution of higher twist effects
to g$_{2}^{n}$, Eq.~\ref{eq:moments} can be solved for the twist-3 
reduced matrix elements $d_{j}^{n}$ at fixed $Q^2$,
\begin{equation}
d_{j}^{n}(Q^{2})=2\int_{0}^{1}x^{j}\left[{\rm{g}}_{1}^{n}(x,Q^{2})
+\left(
\frac{j+1}{j}\right){\rm{g}}_{2}^{n}(x,Q^{2})\right]dx,\;\; j=2,4,...
\label{eq:d2}
\end{equation}
The combination of g$_{1}$ and g$_{2}$ in the above expression
effectively cancels any twist-2 components allowing us to 
look for a net twist-3 contribution to g$_{2}$.
The matrix element was calculated using our g$_{2}(x,Q^{2})$ data and a fit
to our measured g$_{1}(x,Q^{2})$.  Because the integrand in
Eq.~\ref{eq:d2} is purely
twist-3, we assumed the unmeasured region $0.7\leq x \leq 1$ behaves like
$(1-x)^{2}$ as suggested by Brodsky {\em{et al.}}~\cite{Brodsky}, and
fit our last data point to this form to extrapolate to $x=1$.
We neglected any contribution from the region $0\leq x < 0.014$ because 
it is suppressed by the $x^{j}$ term.  
For the $j\!=\!2$ moment, we obtain a value of
$d_{2}^{n}=-0.004\pm 0.038 \pm 0.005$ with an average $Q^{2}$ of $3.6$ 
(GeV/c)$^{2}$.
The contribution from the  high-$x$ extrapolation is much smaller 
than the experimental errors and
does not significantly change the result for the matrix element.

\begin{figure}
{\epsfig{figure=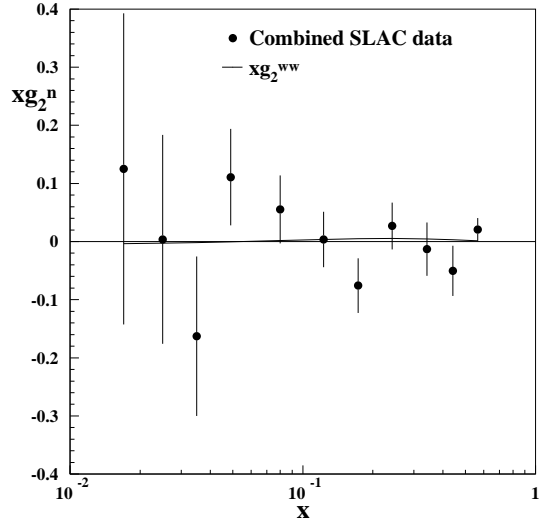,width=7.0cm}}
\caption{The structure function $x$g$_{2}^{n}$  
for SLAC experiments E142, E143 and E154 combined.  Also shown is the 
twist-2 g$_{2}^{WW}$ prediction.  The average $Q^2$ for the combined 
data is $3.0$ (GeV/c)$^2$.  Errors are statistical only.}
\label{fig3}
\end{figure}

Data from SLAC experiments E142~\cite{E142_2}
and E143~\cite{E143g2}
were combined with this experiment to yield a average neutron result 
for g$_{2}$ with an average $Q^{2}$ of $3.0$ (GeV/c)$^{2}$.  
Neutron results were 
extracted from the E143 proton and deuteron data assuming a 5\%
D-state in the deuteron.  
The results are shown in Fig.~\ref{fig3} 
along with the g$_{2}^{WW}$ prediction.  Comparing the combined data
with g$_{2}^{WW}$ gives a $\chi^{2}/$(dof) of $1.01$,
again indicating good agreement with g$_{2}^{WW}$.
Using the combined data, we obtain the result
$d_{2}^{n}=-0.010\pm 0.015$ at an average $Q^{2}$ of $3.0$ (GeV/c)$^{2}$.  
The measured $d_{2}^{n}$ along with various model predictions are
summarized in Table~\ref{tab:d2}, and while the data are consistent
with zero, the precision is insufficient to rule out models which 
contain significant twist-3 contributions.
\begin{table}
\caption{Comparison of experimental and theoretical results for the
reduced twist-3 matrix element $d_{2}^{n}$.}
\begin{tabular}{lcc}
\hline \hline
& $d_{2}^{n}\times 10^{2}$ & $Q^{2}$ (GeV/c)$^{2}$\\ \hline
E154 result & $-0.4\pm 3.8$ & $3.6$\\
SLAC Average & $-1.0\pm 1.5$ & $3.0$\\ \hline
Bag model~\cite{Song} & $-0.253$ & $5.0$\\
Bag model~\cite{E143g2,Stratmann} & $0.03$ & $5.0$\\
Bag model~\cite{Ji} & $0$ & $5.0$\\
QCD sum rule~\cite{Stein} & $-3\pm 1$ & $1.0$\\
QCD sum rule~\cite{BBK} & $-2.7\pm 1.2$ & $1.0$\\
Lattice QCD~\cite{LQCD} & $-0.39\pm 0.27$ & $4.0$\\
\hline \hline
\end{tabular}
\label{tab:d2}
\end{table}

From the OPE it is not possible to obtain an expression for the $j\!=\!0$
moment of g$_{2}$.  However, Burkhardt and Cottingham~\cite{BCsum}
have derived the sum rule $\int_{0}^{1}{\rm{g}}_{2}(x)dx=0$, which is
valid to first order in pQCD~\cite{altarelli}, using dispersion
relations for virtual Compton scattering.  To evaluate the integral, 
the g$_{2}^{WW}$ expression in Eq.~\ref{eq:g2ww} was used to evolve
the twist-2 part of our measured g$_{2}$ to a $Q^{2}$ of $3.6$ 
(GeV/c)$^{2}$ assuming g$_{1}/F_{1}$ is independent of $Q^{2}$
and fitting our g$_{1}^{n}$ data as before.  At large $x$, we see from
Eq.~\ref{eq:g2ww} that g$_{2}^{WW} \approx -$g$_1$ and we therefore assume 
that g$_{2}\propto (1-x)^{3}$ for the extrapolation to $x=1$.  The
result is
$\int_{0.014}^{1}{\rm{g}}_{2}(x)dx=0.19\pm 0.17 \pm 0.02$ with an average
$Q^{2}$ of $3.6$ (GeV/c)$^{2}$.  The $Q^{2}$ evolution and high-$x$
extrapolation do not contribute significantly to the integral
and the uncertainties in these quantites are included in the error.
Combining this result with data from SLAC experiments
E142~\cite{E142_2} and E143~\cite{E143g2} yields a result of
$\int_{0.014}^{1}{\rm{g}}_{2}(x)dx=0.06\pm 0.15$ at an average
$Q^{2}$ of $3.0$ (GeV/c)$^{2}$, which is consistent with the
Burkhardt-Cottingham sum rule.     
However, this does not represent a conclusive test of the sum rule
because the behavior of g$_{2}^{n}$ as $x\rightarrow\!0$ is not known.

In summary, we have presented a new measurement of A$_{2}^{n}$ and g$_{2}^{n}$
in the kinematic range $0.014 \leq x \leq 0.7$ and $1.0 \leq Q^{2}
\leq 17.0$ (GeV/c)$^{2}$.  Our results for A$_{2}^{n}$ are
significantly smaller than the $\sqrt{R}$ positivity limit over most 
of the measured range and data for g$_{2}^{n}$ are generally consistent
with the twist-2 g$_{2}^{WW}$ prediction.  The values obtained for the
twist-3 matrix element $d_{2}^{n}$ from this measurement and the SLAC 
average are also consistent with zero.  However, further measurements 
are needed to make a conclusive statement about the higher twist
content of the nucleon.

We wish to thank the personnel of the SLAC accelerator department for
their efforts which resulted in the successful completion of the E154
experiment.  We would also like to thank J. Ralston for 
useful discussions and guidance.  This work was supported by the
Department of Energy; by
the National Science Foundation; by the Kent State University Research
Council (GGP);  by the Jeffress Memorial Trust (KAG); by the Centre
National de la Recherche Scientifique and the Commissariat a l'Energie
Atomique (French groups); and by the Japanese Ministry of Education,
Science and Culture (Tohoku).\\

\clearpage

\end{document}